\documentclass[12pt]{article}
\usepackage{amssymb,amsmath,epsfig}

\begin{document}

\title{\bf Analysis of $F(R,T)$ Gravity Models Through Energy Conditions}
\author{ M. Sharif$^1$ \thanks{msharif.math@pu.edu.pk}, S. Rani$^1$
\thanks{shamailatoor.math@yahoo.com} and
R. Myrzakulov$^2$ \thanks{rmyrzakulov@csufresno.edu}\\
$^1$Department of Mathematics, University of the Punjab,\\
Quaid-e-Azam Campus, Lahore-54590, Pakistan.
\\
$^2$Eurasian International Center for Theoretical Physics,   \\
Eurasian National University, Astana 010008, Kazakhstan}
\date{}
\maketitle

\begin{abstract}
This paper is devoted to study the energy conditions in $F(R,T)$
gravity for FRW universe with perfect fluid, where $R$ is the Ricci
scalar and $T$ is the torsion scalar. We construct the general
energy conditions in this theory and reduce them in $F(R)$ as well
as $F(T)$ theory of gravity. Further, we assume some viable models
and investigate bounds on their constant parameters to satisfy the
energy condition inequalities. We also plot some of the cases using
present-day values of the cosmological parameters. It is interesting
to mention here that the model $F(R,T)=\mu R+ \nu T$ satisfies the
energy conditions for different ranges of the parameters.
\end{abstract}

\section{Introduction}

Modified theories of gravity have recently attained much attention
to explore the dark energy (DE) and late accelerated expansion of
the universe \cite{1}. Dark energy is almost equally distributed in
the universe and different models have been developed in the last
decade to describe its nature. Dark energy is related with the
modification of Einstein gravity in such a way that it would give
the gravitational alternative to DE. The modified theories of
gravity explain the unification of dark matter and DE, deceleration
to acceleration epochs of the universe, description of hierarchy
problem, dominance of effective DE, which help to solve the
coincidence problem, viability of DE models through energy
conditions and many more \cite{1,2}. The proposed modified theories
of gravity include Gauss-Bonnet gravity with $G$ invariant \cite{3},
$F(R)$ takes Ricci scalar $R$ \cite{4}, $F(T)$ inherits torsion
scalar $T$ \cite{5}, $F(R,T)$ with $T$ as the trace of stress-energy
tensor \cite{6}, $F(R,G)$ carries both $R$ and $G$ \cite{7} etc.
These theories use their corresponding invariant and scalars in
order to meet the acceleration of the expanding universe.

In alternative theories, the $F(R)$ theory has shown significant
progress in cosmology. This theory can directly be achieved by
replacing the Ricci scalar $R$ by $F(R)$ as an arbitrary function in
the Einstein-Hilbert action \cite{1,4}, i.e.,
\begin{equation}\label{a}
S_{EH}=\frac{1}{2\kappa^{2}}\int d^{4}x \sqrt{-g}R\rightarrow
S=\frac{1}{2\kappa^{2}}\int d^{4}x \sqrt{-g}F(R),
\end{equation}
where $\kappa^2=8\pi G,~G$ is the gravitational constant and $g$ is
the determinant of the metric coefficients. This theory uses
Levi-Civita connection having only curvature for its formation and
admits $1/R$ type terms in the Lagrangian to discuss the expanding
universe with acceleration. Following the same strategy, the $F(T)$
theory of gravity is obtained by replacing torsion scalar $T$ with a
general function $F(T)$ in the action of teleparallel equivalent of
general relativity (GR) \cite{5}, i.e.,
\begin{equation}\label{b}
S_{T}=\frac{1}{2\kappa^{2}}\int d^{4}x \sqrt{-g}T\rightarrow
S=\frac{1}{2\kappa^{2}}\int d^{4}x \sqrt{-g}F(T).
\end{equation}
This theory uses Weitzenb\"{o}ck connection which has only torsion
but no curvature. The accelerated expansion of the universe in
$F(T)$ theory is remarked with $T$ formed by the tetrad field.

The unification of $F(R)$ and $F(T)$ gravity theories as $F(R,T)$ is
the prospective and interesting version in modified theories. The
torsion as well as curvature scalar with no matter source leads to
the formation of this unification. There are many extensions of GR
in which torsion effects are included \cite{B}. The $F(R,T)$ theory
of gravity incorporates all the properties of its constituent
gravities. This theory is considered as an important gravitational
theory which represents the evolution of the universe. Myrzakulov
\cite{a} constructed different reductions of $F(R,T)$ gravity and
derived some torsion scalar solutions, in particular, the exact de
Sitter solution by assuming a particular model $F(R,T)=\mu R+\nu T$.
He concluded that these exact analytic solutions describe the
evolving universe in phantom acceleration phase. Chattopadhyay
\cite{b} found a quintom-like behavior and transition from
deceleration to acceleration phase of the universe using the same
model with interacting Ricci DE in this theory. Also, the
statefinder parameters indicate that the model interpolates between
dust and $\Lambda$CDM phases of the universe.

The energy conditions are used to restrict some higher order terms
of GR models or its extensions to make them simpler with viability.
Garc\'{\i}a et al. \cite{c} formulated these energy conditions in
modified Gauss-Bonnet gravity for two DE models and provided the
ranges of some parameters in the model. Santos et al. \cite{d}
examined the bounds on the parameters of two families of $F(R)$
gravity. Liu and Rebou\c{c}as \cite{e} investigated the constraints
on the logarithmic and exponential models in the framework of $F(T)$
gravity. Recently, Alvarenga et al. \cite{f} explored energy
conditions in $F(R,T)$ gravity with $T$ as trace of the
energy-momentum tensor for two specific models. They found that for
some values of the input parameters of the models, the de Sitter and
power-law solutions may be stable with the validity of energy
conditions.

In this paper, we construct the energy conditions on $F(R,T)$
gravity with $T$ as a torsion scalar in FRW spacetime with perfect
fluid. The constraints on the parameters are investigated for the
the general as well as particular cases by taking viable models of
DE. The paper is organized as follows. In next section, we provide
preliminaries related to the energy conditions and $F(R,T)$ gravity.
Section \textbf{3} is devoted to represent the general form of the
energy conditions in $F(R,T)$ gravity and their reduction to $F(R)$
and $F(T)$ theories of gravity. Also, the graphical behavior of the
constraints on the model parameters is shown. The last section
summarizes the results.

\section{Preliminaries}

In this section, we briefly discuss the energy conditions and their
cosmological implications in GR as well as in modified theory of
gravity. The energy-momentum distribution and stress due to matter
or any other non-gravitational field is described by the
energy-momentum tensor $T_{\mu\nu}$. For a realistic matter source
satisfied by all states of matter and non-gravitational fields, it
must satisfy some conditions on this tensor. There are several
different types of these conditions, so called energy conditions
such as averaged energy conditions (depend on the average
stress-energy tensor along a desirable curve) and point-wise energy
conditions (count the stress-energy tensor at a given point in the
space) etc. The standard point-wise energy conditions are the null
(NEC), weak (WEC), strong (SEC) and dominant energy condition (DEC).
Basically these conditions are formulated with the help of
Raychaudhuri equation that describes the behavior of timelike,
spacelike or lightlike curves of a congruence and attractiveness of
the gravity \cite{A}.

The Raychaudhuri equation for timelike and spacelike curves is given
by
\begin{eqnarray}\nonumber
R_{\mu\nu}u^\mu u^\nu+\sigma_{\mu\nu}\sigma^{\mu\nu}
-\omega_{\mu\nu}\omega^{\mu\nu}+\frac{1}{3}\theta^2+\frac{d\theta}{d\pi}=0,\\\nonumber
R_{\mu\nu}k^\mu k^\nu+\sigma_{\mu\nu}\sigma^{\mu\nu}
-\omega_{\mu\nu}\omega^{\mu\nu}+\frac{1}{2}\theta^2+\frac{d\theta}{d\lambda}=0.
\end{eqnarray}
Here $R_{\mu\nu}$ is the Ricci tensor, $u^{\mu}$ and $k^{\mu}$ are
the timelike and lightlike tangent vectors, $\sigma^{\mu\nu}$ and
$\omega^{\mu\nu}$ are the shear and vorticity tensors describing the
distortion of volume and rotation of the curves respectively. The
expansion scalar $\theta$ shows the expansion of volume, while $\pi$
and $\lambda$ are the positive parameters which represent the curved
of the congruence under spacetime manifold. The quadratic terms in
Raychaudhuri equation may be neglected for small distortions of the
volume without rotation, leading to
\begin{equation}\label{2}
\theta=-\pi R_{\mu\nu}u^\mu u^\nu=-\lambda R_{\mu\nu}u^\mu u^\nu.
\end{equation}

The expansion scalar, $\theta<0$, provides the attractiveness
property for any hypersurface of orthogonal congruence
$(\omega_{\mu\nu}=0)$ yielding $R_{\mu\nu}u^{\mu} u^\nu\geq0$ and
$R_{\mu\nu}k^\mu k^\nu\geq0$. In the framework of GR, through the
Einstein field equations, the Ricci tensor is replaced by the
engery-momentum tensor $T_{\mu\nu}$, which leads to point-wise
energy conditions. Taking into account the perfect fluid
($T_{\mu\nu}=(\rho+p)u_{\mu}u_{\nu}+pg_{\mu\nu}$) with energy
density $\rho$ and pressure $p$ in the effective gravitational
field, these energy conditions are classified as follows.

The \textbf{NEC} is the consequence of
$T_{\mu\nu}k^{\mu}k^{\nu}\geq0$, which leads to familiar form
$\rho_{eff}+p_{eff}\geq0$. The NEC implies that density of the
universe falls with its expansion and its violation may yield the
Big Rip of the universe. It ensures the validity of second law of
black hole thermodynamics. The positivity condition for timelike
vector $u^{\mu}$ results the \textbf{SEC} which reduces to
$\rho_{eff}+p_{eff}\geq0,~\rho_{eff}+3p_{eff}\geq0$ in the effective
field. The violation of this condition represents the accelerated
expansion of the universe. The \textbf{WEC} requires the positivity
of the energy density for any observer at any point, i.e.,
$\rho_{eff}\geq0,~\rho_{eff}+p_{eff}\geq0$ in addition to NEC. The
Hawking-Penrose singularity theorems require the validity of SEC and
WEC. The condition on energy that it must not flow faster than light
$(p\leq\rho)$ yields one inequality of \textbf{DEC} whose complete
set is $\rho_{eff}\geq0,~\rho_{eff}\pm p_{eff}\geq0$. The WEC and
NEC are the most important of all energy conditions as their
violation lead to the violation of other energy conditions.

This procedure is trivially true for $F(R)$ gravity. However, for
any other theory, the replacement of the energy-momentum tensor
should be taken under consideration. It may not affect the procedure
if NEC is satisfied for any matter and is maintaining by the
physical motivation of geodesic congruences. The Raychaudhuri
equation and attractiveness property hold for any gravitational
theory, and hence can be used to check the energy conditions of the
recently developed modified theories of gravity.

There is an equivalent role treated by both curvature and torsion
scalars in the fundamental framework. It should not be surprising
that such a comparability between curvature and torsion should be
found in the construction of the gravitational Lagrangian of the
theory. Consider the flat, homogenous and isotropic FRW universe as
\begin{equation}\label{38}
ds^{2}=-dt^{2}+a^{2}(t)(dx^{2}+dy^{2}+dz^{2}),
\end{equation}
where $a$ is the time dependent scale factor. The corresponding
Lagrangian of $M_{37}$ - $F(R,T)$ gravity model is given by
\cite{a,b}
\begin{eqnarray}\nonumber
L_{37}&=&a^{3}[F-(T-v)F_{T}-(R-u)F_{R}+L_{m}]-6a{\dot{a}}^{2}(F_R+F_T)
\\\label{5}
&-&6a^{2}\dot{a}(\dot{R}F_{RR}+\dot{T}F_{RT}).
\end{eqnarray}
Here $L_m$ is the matter Lagrangian, the subscripts represent the
first and second order derivatives of $F$ with respect to the
corresponding scalar and dot represents the time derivative. The
functions $u$ and $v$ are generally defined as
$u=u(t,a,\dot{a},\ddot{a},\dddot{a},...)$ and
$v=v(t,a,\dot{a},\ddot{a},\dddot{a},...)$, which are related with
the geometry of the spacetime. The Ricci and torsion scalars are
defined as
\begin{equation}\label{39}
R=u+g^{\mu\nu}R_{\mu\nu},\quad
T=v-S_{\rho}~^{\mu\nu}T^{\rho}~_{\mu\nu},
\end{equation}
where $g^{\mu\nu}$ is the metric tensor and $R_{\mu\nu},~
S_{\rho}~^{\mu\nu},~T^{\lambda}~_{\mu\nu}$ are the Ricci,
antisymmetric and torsion tensors respectively, given by
\begin{eqnarray}\label{40}
R_{\mu\nu}&=&\partial_{\lambda}\Gamma^{\lambda}_{\mu\nu}-\partial_{\mu}\Gamma^{\lambda}_{\lambda
\nu}+\Gamma^{\lambda}_{\mu\nu}\Gamma^{\rho}_{\rho\lambda}-\Gamma^{\lambda}_{\nu\rho}
\Gamma^{\rho}_{\mu\lambda},\\\label{41}S_{\rho}~^{\mu\nu}&=&-\frac{1}{4}(T^{\mu\nu}~_{\rho}-
T^{\nu\mu}~_{\rho}-T_{\rho}~^{\mu\nu})
+\frac{1}{2}(\delta^{\mu}_{\rho}T^{\theta\nu}~_{\theta}-\delta^{\nu}_{\rho}T^{\theta\mu}~_{\theta}),\\
\label{42}T^{\lambda}~_{\mu\nu}&=&\Gamma^{\lambda}~_{\nu\mu}-
\Gamma^{\lambda}~_{\mu\nu}=h^{\lambda}_{a}
(\partial_{\nu}h^{a}_{\mu}-\partial_{\mu}h^{a}_{\nu}).
\end{eqnarray}
Here $h^{a}_{\mu}$ are the tetrad components and
$\Gamma^{\lambda}_{\mu\nu}=\frac{1}{2}g^{\lambda
\rho}(\partial_{\mu} g_{\nu \rho}+\partial_{\nu} g_{\rho
\mu}-\partial_{\rho} g_{\mu \nu})$ are the Christoffel symbols with
Greek alphabets $(\mu,\nu,\rho,...=0,1,2,3)$ denote spacetime
components while the Latin alphabets $(a,b,c,...=0,1,2,3)$ are used
to describe components of tangent space..

Using Eq.(\ref{38}) and tetrad $h^{a}_{\mu}=diag(1,a,a,a)$, the
Ricci and torsion scalars become $R=u+6(\dot{H}+2H^2)$ and
$T=v-6H^2$ with Hubble parameter $H=\dot{a}/a$. Taking $a,R$ and $T$
as the generalized coordinates of the configuration space, the total
energy (Hamiltonian) corresponding to the Lagrangian (\ref{5})
yields the following field equations (assuming $\kappa^{2}=1$)
\begin{eqnarray}\nonumber
&&6a^2\dot{a}\dot{R}F_{RR}-(6a^2\ddot{a}+a^3\dot{a}\frac{\partial
u}{\partial
\dot{a}})F_R+6a^{2}\dot{a}\dot{T}F_{RT}+(12a{\dot{a}}^{2}
-a^{3}\dot{a}\frac{\partial v}{\partial
\dot{a}})F_T\\\label{3}&&+a^3F=2a^3\rho_m,\\\nonumber
&&-6a^{2}{\dot{R}}^{2}F_{RRR}-(12a\dot{a}\dot{R}+6a^{2}\ddot{R}-a^{3}\dot{R}\frac{\partial
u}{\partial \dot{a}})F_{RR}+(12{\dot{a}}^{2}+6a\ddot{a}\\\nonumber
&&+3a^{2}\dot{a}\frac{\partial u}{\partial
\dot{a}}+a^{3}\frac{\partial}{\partial t}(\frac{\partial u}{\partial
\dot{a}})-a^{3}\frac{\partial u}{\partial
a})F_{R}-(12a\dot{a}\dot{T}-a^{3}\dot{T}\frac{\partial v}{\partial
\dot{a}})F_{TT}\\\nonumber
&&-(24{\dot{a}}^{2}+12a\ddot{a}-3a^{2}\dot{a}\frac{\partial
v}{\partial \dot{a}}-a^{3}\frac{\partial}{\partial t}(\frac{\partial
v}{\partial \dot{a}})+a^{3}\frac{\partial v}{\partial
a})F_{T}-12a^{2}\dot{R}\dot{T}F_{RRT}\\\nonumber
&&-6a^{2}{\dot{T}}^{2}F_{RTT}
-(12a\dot{a}\dot{T}+12a\dot{a}\dot{R}+6a^{2}\ddot{T}-a^{3}\dot{R}\frac{\partial
v}{\partial \dot{a}}-a^{3}\dot{T}\frac{\partial u}{\partial
\dot{a}})F_{RT}\\\label{4}&&-3a^{2}F=6a^{2}p_m.
\end{eqnarray}

\section{Energy Conditions of $F(R,T)$ Gravity}

In this section, we construct the energy conditions of $M_{37}$ -
$F(R,T)$ gravity model. The $M_{37}$ - model admits some important
features from the physically and geometrically viewpoints. We assume
the functions $u$ and $v$ in power-law form \cite{b} as $u=\alpha
a^n$ and $v=\beta a^m$ where $\alpha,~\beta$ are non-zero real
constants and $m,~n$ are positive integers. The Ricci and torsion
scalars with their derivatives can be expressed in terms of
deceleration $(q)$, jerk $(j)$ and snap $(s)$ parameters \cite{g} as
\begin{eqnarray}\nonumber
&&R=\alpha a^{n}+6H^2(1-q),\quad \dot{R}=\alpha
na^{n}H+6H^{3}(j-q-2),\\\label{7} &&\ddot{R}=\alpha
na^{n}(n-q-1)H^2+6H^{4}(4q^2+15q+2j+s+9),\\\nonumber &&T=\beta
a^{m}-6H^2,\quad \dot{T}=\beta ma^{m}H+12H^{2}(1+q),\\\label{8}
&&\ddot{T}=\beta ma^{m}(m-q-1)-12H^{4}(q^2+5q+j+3),
\end{eqnarray}
where $q=-\frac{1}{H^2}\frac{\ddot{a}}{a},~
j=\frac{1}{H^3}\frac{\dddot{a}}{a},~
s=\frac{1}{H^4}\frac{\ddddot{a}}{a}$. The present-day values of $H$
and negative $q$ describe the expansion rate of the accelerating
universe, while $j$ classifies different DE models representing
acceleration of the universe. Equations (\ref{3}) and (\ref{4}) can
be written as the effective gravitational field equations in the
following form
\begin{equation}\label{6}
3H^2=\rho_{eff},\quad -(2\dot{H}+3H^2)=p_{eff},
\end{equation}
where
\begin{eqnarray}\nonumber
\rho_{eff}&=&\rho_m-3H^{2}(\alpha
na^{n}+6H^{2}(j-q-2))F_{RR}-(3H^2q+\frac{\alpha
na^{n}}{2q})F_{R}\\\nonumber&-&3H^2(\beta
ma^{m}+12H^{2}(1+q))F_{RT}-(6H^{2}+\frac{\beta
ma^m}{2q})F_{T}+3H^2\\\label{9}&-&\frac{F}{2},\\\nonumber
p_{eff}&=&p_m+(2q-1)H^{2}+H^{2}(\alpha
na^{n}+6H^{2}(j-q-2))^{2}F_{RRR}\\\nonumber&+&[\alpha
na^{n}H^{2}(1-q+n)+6H^{4}(4q^{2}+4j+13q+s+5)\\\nonumber&+&\frac{\alpha
na^{n}}{6q}(\alpha
na^{n}+6H^{2}(j-q-2))]F_{RR}-[(2-q)H^{2}-\frac{\alpha
na^{n}}{6}\\\nonumber&\times&(\frac{n+2}{q}+1)]F_{R}+(2H^{2}+\frac{\beta
ma^{m}}{6q})(\beta
ma^{m}+12H^{2}(1+q))F_{TT}\\\nonumber&+&[2(2-q)H^{2}+\frac{\beta
ma^{m}}{6}(\frac{m+2}{q}+1)]F_{T}+2H^{2}(\alpha
na^{n}\\\nonumber&+&6H^{2}(j-q-2)(\beta
ma^{m}+12H^{2}(1+q))F_{RRT}+H^{2}(\beta
ma^{m}\\\nonumber&+&H^{2}(1+q))^{2}F_{RTT}+[\beta
ma^{m}(m-q+1)H^{2}+2\alpha na^{n}H^{2}\\\nonumber
&-&12H^{4}(q^{2}+4q+3)+\frac{\beta ma^{m}}{6q}(\alpha
na^{n}+6H^{2}(j-q-2))\\\label{10}&+&\frac{\alpha na^{n}}{6q}(\beta
ma^{m}+12H^{2}(1+q))]F_{RT}+\frac{F}{2}.
\end{eqnarray}

Now we discuss energy conditions for a particular model \cite{a,b}
\begin{equation}\label{31}
F(R,T)=\mu R+\nu T,
\end{equation}
where $\mu$ and $\nu$ are non-zero real constants. This model
describes the accelerated expansion of the universe in phantom era
by taking power-law form for the functions $u$ and $v$. Inserting
model (\ref{31}) in Eqs.(\ref{9}) and (\ref{10}), we obtain the
effective field equations as
\begin{eqnarray}\nonumber
\rho_{eff}&=&\rho_m+3H^{2}-\mu(3H^{2}q+\frac{\alpha
na^{n}}{2q})-\nu(6H^{2}+\frac{\beta
ma^m}{2q})\\\label{32}&-&\frac{\mu R+\nu T}{2}, \\\nonumber
p_{eff}&=&p_m+(2q-1)H^{2}-\mu[(2-q)H^{2}-\frac{\alpha
na^{n}}{6}(\frac{n+2}{q}+1)]\\\label{33}&+&\nu[2(2-q)H^{2}+\frac{\beta
ma^{m}}{6}(\frac{m+2}{q}+1)]+\frac{\mu R+\nu T}{2}.
\end{eqnarray}
The expressions for NEC, WEC, SEC and DEC using Eqs.(\ref{32}) and
(\ref{33}) take the form
\begin{eqnarray}\nonumber
&&\textbf{NEC}:\quad\rho_{m0}+p_{m0}+2(1+q_{0})H_{0}^{2}-\mu[2(1+q_{0})H_{0}^{2}+\frac{\alpha
na_{0}^{n}}{2}(\frac{1-n-q_{0}}{3q_{0}})]\\\label{34}&&-\nu[2(1+q_{0})H_{0}^{2}+\frac{\beta
ma_{0}^{m}}{2}(\frac{1-m-q_{0}}{q_{0}})]\geq0,
\end{eqnarray}
\begin{eqnarray}\nonumber
&&\textbf{WEC}:\quad\rho_{m0}+3H_{0}^{2}-\mu(3H_{0}^{2}q_{0}+\frac{\alpha
na_{0}^{n}}{2q_{0}})-\nu(6H_{0}^{2}+\frac{\beta
ma_{0}^m}{2q_{0}})\\\label{35}&&-\frac{\mu}{2}(\alpha
a_{0}^n+6H_{0}^{2}(1-q_{0}))-\frac{\nu}{2}(\beta
a_{0}^m-6H_{0}^{2})\geq0,\\\nonumber
&&\textbf{SEC}:\quad\rho_{m0}+3p_{m0}+6H_{0}^{2}q_{0}-\mu(6H_{0}^{2}-\frac{\alpha
na_{0}^{n}}{2}(\frac{1+n+q_{0}}{q_{0}}))\\\nonumber
&&+\nu(6(1-q_{0})H_{0}^{2}+\frac{\beta
ma_{0}^m}{2}(\frac{1+m+q_{0}}{q_{0}}))+\frac{\mu}{2}(\alpha
a_{0}^n+6H_{0}^{2}(1-q_{0}))\\\label{36}&&+\frac{\nu}{2}(\beta
a_{0}^m-6H_{0}^{2})\geq0,\\\nonumber
&&\textbf{DEC}:\quad\rho_{m0}-p_{m0}+2(2-q_{0})H_{0}^{2}-\mu(2(2q_{0}-1)H_{0}^{2}+\frac{\alpha
na_{0}^{n}}{2}(\frac{n+q_{0}+5}{3q_{0}}))\\\nonumber
&&-\nu(2(5-q_{0})H_{0}^{2}+\frac{\beta
ma_{0}^m}{2}(\frac{m+q_{0}+5}{3q_{0}})-\frac{\mu}{2}(\alpha
a_{0}^n+6H_{0}^{2}(1-q_{0}))\\\label{37}&&-\frac{\nu}{2}(\beta
a_{0}^m-6H_{0}^{2})\geq0,
\end{eqnarray}
where $\rho_{eff}+p_{eff}\geq0$ for NEC, $\rho_{eff}\geq0$ for WEC,
$\rho_{eff}+3p_{eff}\geq0$ for SEC, $\rho_{eff}-p_{eff}\geq0$ for
DEC are formulated and the subscript $0$ denotes the present-day
values of the corresponding parameters. In particular, we consider
the following present-day values \cite{g}:
$H_{0}=0.718,~q_{0}=-0.64,~j_{0}=1.02,~s_{0}=-0.39$ and $a_{0}=1$ to
discuss the energy conditions. Also, we assume $\rho_{m0}=0=p_{m0}$
so that if the vacuum $F(R,T)$ model satisfies the WEC then one can
always add a positive energy density or pressure from matter and
radiation to any model satisfying the WEC.
\begin{figure}
\epsfig{file=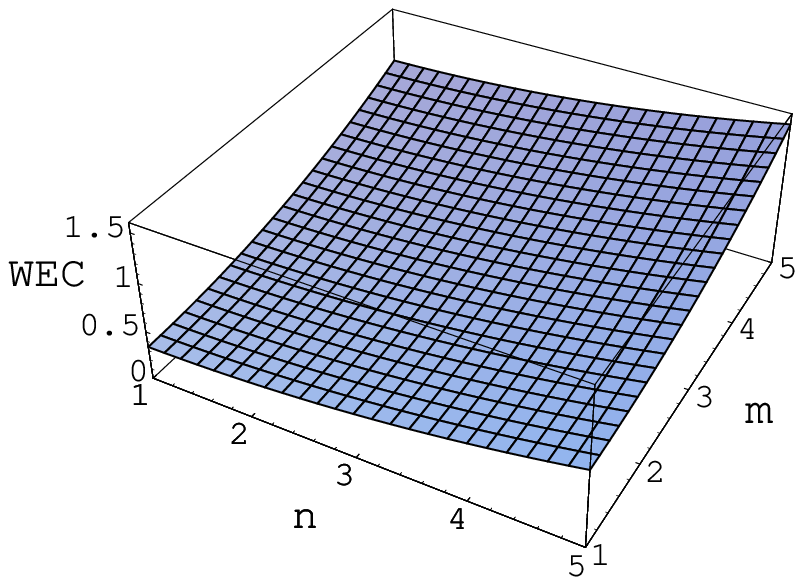, width=0.5\linewidth}\epsfig{file=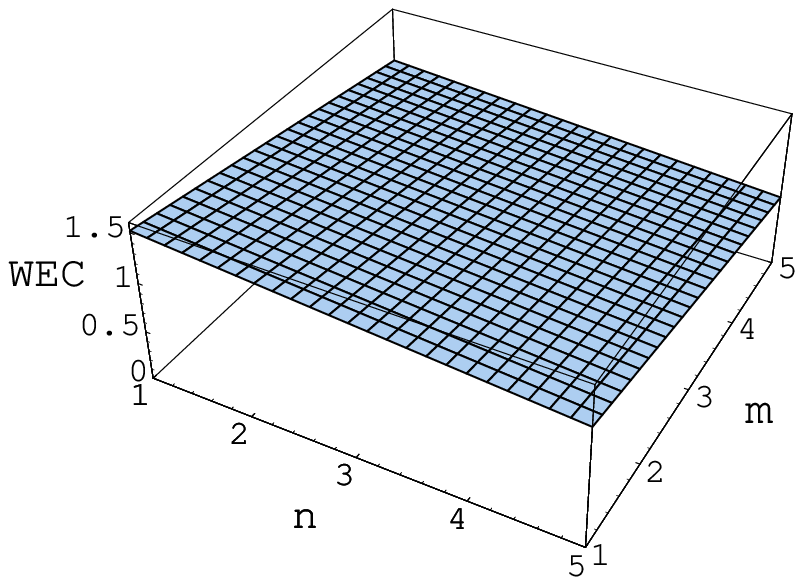,
width=0.5\linewidth}\caption{Plot of WEC versus $n$ and $m$. The
left graph corresponds to $\rho_{eff}+p_{eff}\geq0$ and the right
graph represents $\rho_{eff}\geq0$ for $\alpha =-5,~\beta =3.08,~\mu
=0.02$ and $\nu =-0.05$.}
\end{figure}

The inequalities (\ref{34}) and (\ref{35}) contain six constants
$\alpha,~\beta,~\mu,~\nu,~m$ and $n$. We impose constraints on $\mu$
and $\nu$ with the signs of $\alpha$ and $\beta$ and draw the graph
against $m$ and $n$. The WEC is satisfied for
$-0.1\leq(\mu,~\nu)\leq0.1$ with $\alpha <0$ and $\beta
>0$ or $\alpha >0$ and $\beta <0$. Figure \textbf{1} shows the
corresponding plot for some specific values from these ranges. The
general model (\ref{31}) satisfies the WEC inequalities for the
particular values of the involved parameters. Such types of
constraints on the model parameters are widely constructed in the
literature \cite{c}-\cite{f,m}.

In the following, we present some special reductions of $M_{37}$ -
model and discuss their energy conditions (particularly WEC and NEC)
using some viable models.

\subsection{The $M_{44}$ - Model}

Let the function $F(R,T)$ be independent of the torsion scalar,
i.e., $F(R,T)=F(R)$. It is the $M_{44}$ - model whose Lagrangian
(\ref{5}) takes the form
\begin{equation}\label{11}
L_{44}=a^{3}[F-(R-u)F_{R}+L_{m}]-6a{\dot{a}}^{2}F_R
-6a^{2}\dot{a}\dot{R}F_{RR}.
\end{equation}
The effective field equations (\ref{9}) and (\ref{10}) reduce to
\begin{eqnarray}\nonumber
\rho_{eff}&=&\rho_m-3H^{2}(\alpha
na^{n}+6H^{2}(j-q-2))F_{RR}-(3H^2q+\frac{\alpha
na^{n}}{2q})F_{R}\\\label{12}&+&3H^2-\frac{F}{2},\\\nonumber
p_{eff}&=&p_m+(2q-1)H^{2}+H^{2}(\alpha
na^{n}+6H^{2}(j-q-2))^{2}F_{RRR}\\\nonumber&+&[\alpha
na^{n}H^{2}(1-q+n)+6H^{4}(4q^{2}+4j+13q+s+5)\\\nonumber&+&\frac{\alpha
na^{n}}{6q}(\alpha
na^{n}+6H^{2}(j-q-2))]F_{RR}-[(2-q)H^{2}\\\label{13}&-&\frac{\alpha
na^{n}}{6}(\frac{n+2}{q}+1)]F_{R}+\frac{F}{2}.
\end{eqnarray}
The energy conditions will become
\begin{eqnarray}\nonumber
&&\textbf{NEC}:\quad\rho_{m}+p_{m}+2(q+1)H^{2}+H^{2}(\alpha
na^{n}+6H^{2}(j-q-2))^{2}F_{RRR}\\\nonumber&&-[\alpha
na^{n}(q+n+2)H^{2}-6H^{4}(4q^{2}+j+16q+s+11)-\frac{\alpha
na^{n}}{6q}(\alpha
na^{n}\\\label{14}&&+6H^{2}(j-q-2))]F_{RR}-[2(1+q)H^{2}+\frac{\alpha
na^{n}}{2}(\frac{1-n-q}{3q})]F_{R}\geq0,\\\nonumber
&&\textbf{WEC}:\quad\rho_m-3H^{2}(\alpha
na^{n}+6H^{2}(j-q-2))F_{RR}-(3H^2q+\frac{\alpha
na^{n}}{2q})F_{R}\\\label{15}&&+3H^2-\frac{F}{2}\geq0,\\\nonumber
&&\textbf{SEC}:\quad\rho_{m}+3p_{m}+6H^{2}+3H^{2}(\alpha
na^{n}+6H^{2}(j-q-2))^{2}F_{RRR}\\\nonumber
&&-3[\alpha
na^{n}(2-q-n)H^{2}+6H^{4}(4q^{2}+5j+12q+s+3)+\frac{\alpha
na^{n}}{6q}(\alpha
na^{n}\\\label{16}
&&+6H^{2}(j-q-2))]F_{RR}-[6H^{2}-\frac{\alpha
na^{n}}{2}(\frac{1+n+q}{q})]F_{R}+F\geq0,\\\nonumber
\end{eqnarray}
\begin{eqnarray}\nonumber
&&\textbf{DEC}:\quad\rho_{m}-p_{m}+2(2-q)H^{2}-H^{2}(\alpha
na^{n}+6H^{2}(j-q-2))^{2}F_{RRR}\\\nonumber&&-[\alpha
na^{n}(4-q+n)H^{2}+6H^{4}(4q^{2}+7j+10q+s-1)+\frac{\alpha
na^{n}}{6q}\\\nonumber&&\times(\alpha
na^{n}+6H^{2}(j-q-2))]F_{RR}-[2(2q-1)H^{2}+\frac{\alpha
na^{n}}{2}(\frac{5+n+q}{3q})]F_{R}\\\label{17}&&-F\geq0.
\end{eqnarray}
These are the general conditions of the $M_{44}$ - model which can
be used to evaluate the energy conditions for any viable model. It
is noted that for $u=0$ the inequalities (\ref{14})-(\ref{17})
become equivalent to the standard energy conditions in $F(R)$
gravity \cite{d}.

We assume here an arbitrary viable $F(R)$ model \cite{4,h} given by
\begin{equation}\label{18}
F(R)=R-\frac{b^2}{3R},
\end{equation}
where $b$ is a positive constant and factor $3$ is used only to
simplify the equations. This model belongs to the family $R-1/R$
which is referred to the cosmic expansion of the universe. Also, it
is consistent with the X-rays galaxy cluster distance data \cite{j},
which leads to the deceleration and snap parameters consistent with
three data sets \cite{4,k}. Inserting Eq.(\ref{18}) in (\ref{14})
and (\ref{15}), the expressions for WEC (contains NEC) become
\begin{eqnarray}\nonumber
&&\rho_{m0}+p_{m0}+2(q_{0}+1)H_{0}^{2}+\frac{2b^{2}H_{0}^{2}(\alpha
na_{0}^{n}+6H_{0}^{2}(j_{0}-q_{0}-2))^{2}}{(\alpha
a_{0}^{n}+6H_{0}^{2}(1-q_{0}))^{4}}\\\nonumber&&+\frac{2b^{2}}{3(\alpha
a_{0}^{n}+6H_{0}^{2}(1-q_{0}))^{3}}[\alpha
na_{0}^{n}(q_{0}+n+2)H_{0}^{2}-6H_{0}^{4}(4q_{0}^{2}+j_{0}+16q_{0}\\\nonumber
&&+s_{0}+11)-\frac{\alpha na_{0}^{n}}{6q_{0}}(\alpha
na_{0}^{n}+6H_{0}^{2}(j_{0}-q_{0}-2))]-[2(1+q_{0})H_{0}^{2}\\\label{19}&&+\frac{\alpha
na_{0}^{n}}{2}(\frac{1-n-q_{0}}{3q_{0}})](1+\frac{b^{2}}{3(\alpha
a_{0}^{n}+6H_{0}^{2}(1-q_{0}))^{2}})\geq0,\\\nonumber
&&\rho_{m0}+\frac{2b^{2}H_{0}^{2}}{(\alpha
a_{0}^{n}+6H_{0}^{2}(1-q_{0}))^{3}}(\alpha
na_{0}^{n}+6H_{0}^{2}(j_{0}-q_{0}-2))\\\nonumber
&&-(3H_{0}^2q_{0}+\frac{\alpha
na_{0}^{n}}{2q_{0}})(1+\frac{b^{2}}{3(\alpha
a_{0}^{n}+6H_{0}^{2}(1-q_{0}))^{2}})+3H_{0}^2\\\label{20}&&-\frac{1}{2}(\alpha
a_{0}^{n}+6H_{0}^{2}(1-q_{0})-\frac{b^{2}}{3(\alpha
a_{0}^{n}+6H_{0}^{2}(1-q_{0}))})\geq0,
\end{eqnarray}
\begin{figure}
\epsfig{file=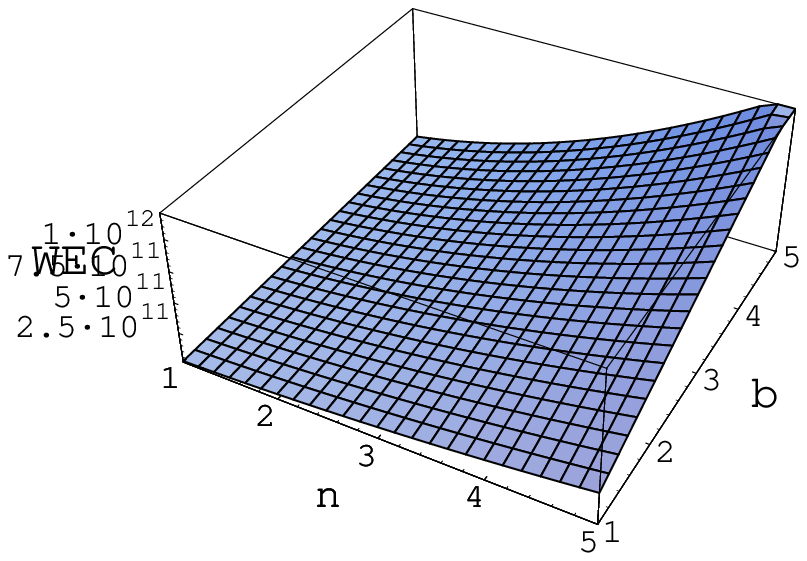, width=0.5\linewidth}\epsfig{file=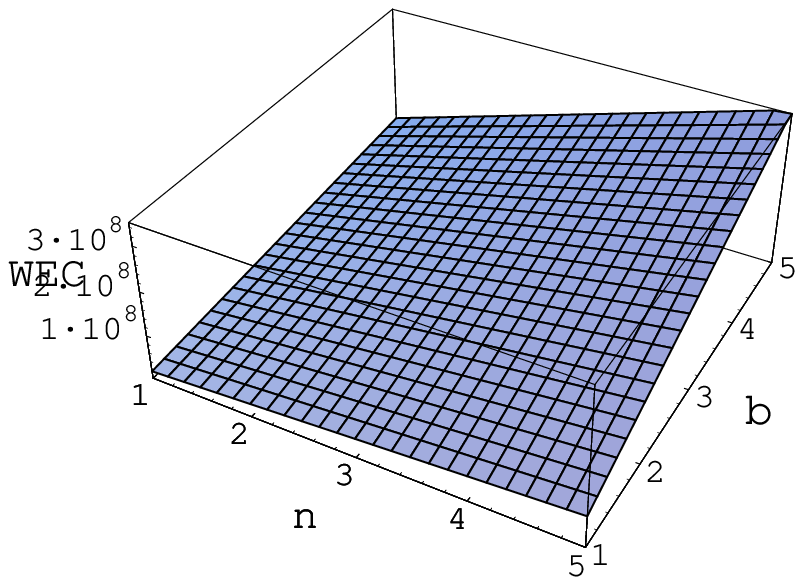,
width=0.5\linewidth}\caption{Plot of WEC versus $n$ and $b$. The
left graph corresponds to $\rho_{eff}+p_{eff}\geq0$ and the right
graph represents $\rho_{eff}\geq0$ for $\alpha =-5.08$.}
\end{figure}

We take here the same assumptions on cosmological parameters as for
the $M_{37}$ - $F(R,T)$ model. Inequalities (\ref{19}) and
(\ref{20}) contain three constants $n,~b$ and $\alpha$. We plot
these conditions against $n$ and $b$ by fixing the range of $\alpha$
as shown in Figure \textbf{2}. For the scales of $n$ and $b$, the
value of $\alpha$ must lie in the range $[-6.1,-5.08]$ to satisfy
these inequalities. Beyond this range, the WEC does not satisfy for
$\alpha>-5.08$ whereas for $\alpha<-6.1$, the inequalities are
supported by higher values of $n$ and $b$ upto $n,b>>70$. Such type
of constraints on the parameters for the validity of the
inequalities have also been constructed in modified theories of
gravity \cite{c}-\cite{f}.

\subsection{The $M_{45}$ - Model}

If the function $F(R,T)$ is independent of $R$, then the model
becomes $M_{45}$ keeping only the dynamics of torsion scalar. In
this case, the Lagrangian (\ref{5}) becomes
\begin{equation}\label{21}
L_{45}=a^{3}[F-(T-v)F_{T}+L_{m}]-6a{\dot{a}}^{2}F_T,
\end{equation}
yielding
\begin{eqnarray}\label{22}
\rho_{eff}&=&\rho_m-(6H^{2}+\frac{\beta
ma^m}{2q})F_{T}+3H^2-\frac{F}{2},\\\nonumber
\end{eqnarray}
\begin{eqnarray}\nonumber
p_{eff}&=&p_m+(2q-1)H^{2}+(2H^{2}+\frac{\beta ma^{m}}{6q})(\beta
ma^{m}+12H^{2}(1+q))F_{TT}\\\label{23}&+&[2(2-q)H^{2}+\frac{\beta
ma^{m}}{6}(\frac{m+2}{q}+1)]F_{T}+\frac{F}{2}.
\end{eqnarray}
Using the above effective field equations, the general form of the
energy conditions are given as follows
\begin{eqnarray}\nonumber
&&\textbf{NEC}:\quad\rho_{m}+p_{m}+2(q+1)H^{2}+(2H^{2}+\frac{\beta
ma^{m}}{6q})(\beta
ma^{m}\\\label{24}&&+12H^{2}(1+q))F_{TT}-[2(1+q)H^{2}+\frac{\beta
ma^{m}}{2}(\frac{1-m-q}{3q})]F_{T}\geq0,\\\label{25}
&&\textbf{WEC}:\quad\rho_m-(6H^{2}+\frac{\beta
ma^m}{2q})F_{T}+3H^2-\frac{F}{2}\geq0,\\\nonumber
&&\textbf{SEC}:\quad\rho_{m}+3p_m+6H^{2}q+3(2H^{2}+\frac{\beta
ma^{m}}{6q})(\beta
ma^{m}+12H^{2}\\\label{26}&&\times(1+q))F_{TT}+[6(1-q)H^{2}+\frac{\beta
ma^{m}}{2}(\frac{1+m+q}{q})]F_{T}+F\geq0,\\\nonumber
&&\textbf{DEC}:\quad\rho_m-p_m+2(2-q)H^{2}-(2H^{2}+\frac{\beta
ma^{m}}{6q})(\beta
ma^{m}+12H^{2}\\\label{27}&&\times(1+q))F_{TT}-[2(5-q)H^{2}+\frac{\beta
ma^{m}}{2}(\frac{m+3q+5}{q})]F_{T}-F\geq0.
\end{eqnarray}
These conditions reduce to the well-known energy conditions of
$F(T)$ gravity in the limit $v=0$ \cite{e}.

Now we evaluate these energy conditions for the following power-law
model \cite{l}
\begin{equation}\label{28}
F(T)=\epsilon ~T^{\delta},
\end{equation}
where $\delta $ is non-zero integer and $\epsilon$ is non-zero real
constant. For $\delta =1= \epsilon$, the model corresponds to
teleparallel gravity. Also, this model helps to remove four types of
singularities emerging in the late-time accelerated era with
specific conditions $\epsilon \neq0$ and $\delta <0$. We check
energy conditions for this model by taking some assumptions on these
parameters. Applying the present-day values of cosmological
parameters and inserting model (\ref{28}) in Eqs.(\ref{24}) and
(\ref{25}), the WEC is obtained as
\begin{eqnarray}\nonumber
&&\rho_{m0}+p_{m0}+2(q_{0}+1)H_{0}^{2}+\epsilon \delta
(\delta-1)(-6H_{0}^{2})^{\delta-2}(2H_{0}^{2}+\frac{\beta
ma_{0}^{m}}{6q_{0}})\\\nonumber&&\times(\beta
ma_{0}^{m}+12H_{0}^{2}(1+q_{0}))-\epsilon \delta
(-6H_{0}^{2})^{\delta-1}[2(1+q_{0})H_{0}^{2}\\\label{29}&&+\frac{\beta
ma_{0}^{m}}{2}(\frac{1-m-q_{0}}{3q_{0}})]\geq0,\\\label{30}&&\rho_{m0}-\epsilon
\delta (-6H_{0}^{2})^{\delta-1}(6H_{0}^{2}+\frac{\beta
ma_{0}^m}{2q_{0}})+3H_{0}^2-\frac{1}{2}\epsilon
(-6H_{0}^{2})^{\delta}\geq0.
\end{eqnarray}
\begin{figure}
\epsfig{file=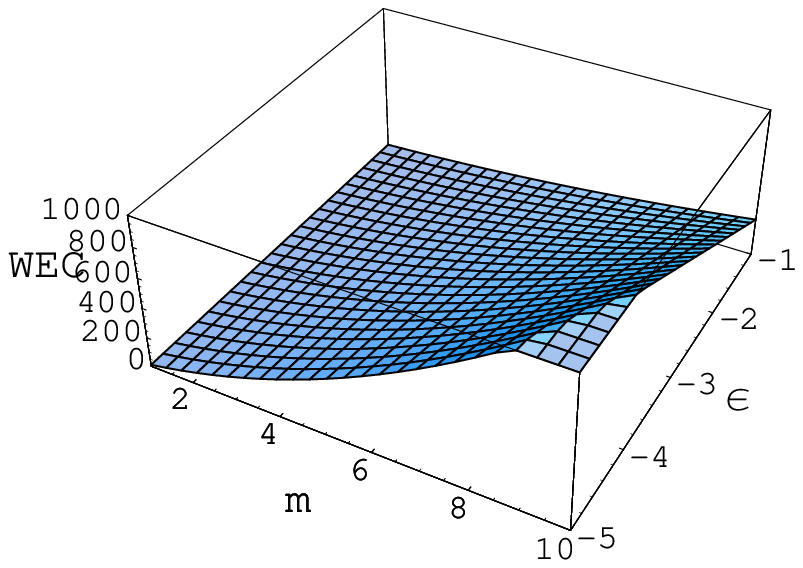, width=0.5\linewidth}\epsfig{file=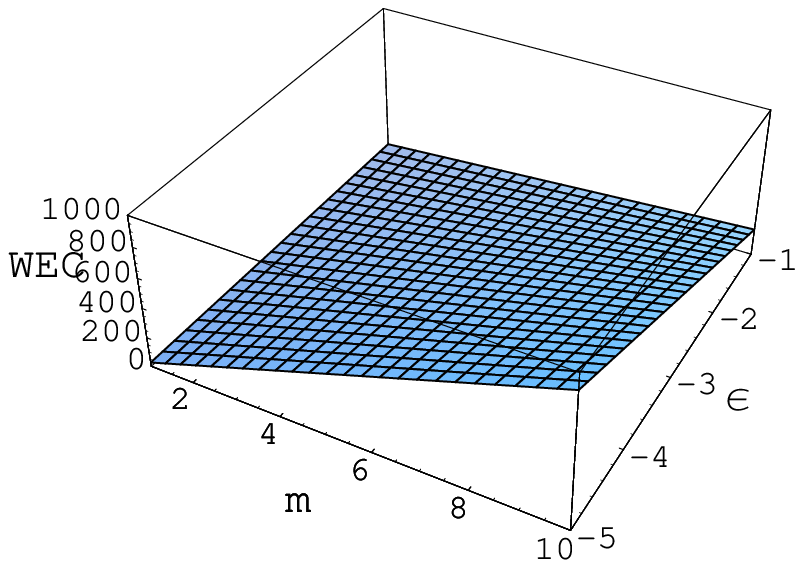,
width=0.5\linewidth}\caption{Plot of WEC versus $m$ and $\epsilon$.
The left graph corresponds to $\rho_{eff}+p_{eff}\geq0$ and the
right graph represents $\rho_{eff}\geq0$ for $\beta =4$ and $\delta
=2$.}
\end{figure}

Taking the present-day values of cosmological parameters and the
condition of vacuum $F(T)$ model similar to the $M_{37}$ - model, we
plot the WEC versus $m$ and $\epsilon$. We have to restrict here one
more constant as compared to the $F(R)$ model. Thus for the model
(\ref{28}), the WEC is valid for $\beta\gg0$ with the following
cases:
\begin{itemize}
\item When $\delta$ is positive,\\
(i) for even $\delta$, it must
take $\epsilon<0$,\\
(ii) if $\delta$ is odd, $\epsilon>0$ is applied.
\item When $\delta$ is negative,\\(i) for even $\delta$,
$\epsilon<0$ with $\delta \leq -6$ whereas $\epsilon>0$ with $\delta
\leq -10$,\\(ii) if $\delta$ is odd, then the ranges for the
parameters are $\epsilon<0$ with $\delta \leq -11$ whereas
$\epsilon>0$ with $\delta \leq -5$.
\end{itemize}
Figure \textbf{3} represents the plot of the first case of positive
$\delta$. The negative $\delta$ constraints satisfying the energy
conditions make the model more reliable to discuss the accelerated
expansion of the universe. In \cite{e}, the WEC is fulfilled for the
specific range of single parameter of the exponential and
logarithmic $F(T)$ models.

\section{Summary}

The energy conditions in the cosmological modeling play an important
role in interpreting different aspects of the universe including the
current accelerated expansion of the universe and singularity
theorems. These conditions are very useful in order to constraint
different constant parameters of the model to be viable in the
underlying framework, in particular NEC and WEC are very important.
In this paper, we use this phenomenon in a recently developed
$F(R,T)$ modified theory of gravity for FRW universe model. Here $T$
is the torsion scalar involved independently in the Lagrangian and
plays its role with the Ricci scalar $R$. We have constructed
expressions of energy conditions for $F(R,T)$ gravity as well as
some special cases with viable models. The ranges of the parameters
are investigated for which the energy conditions hold. Also, we give
graphical representations of some of the parameters for the WEC. The
results of the paper are summarized as follows.

Firstly, we have constructed the inequalities for the energy
conditions of $M_{37}$ - $F(R,T)$ gravity for a particular model
$F(R,T)=\mu R+\nu T$. This model shows consistent results for
$-0.1\leq(\mu,~\nu)\leq0.1$ with different signs of $\alpha$ and
$\beta$. Also, this model represents the accelerated expansion of
the universe \cite{a} and satisfies the energy conditions with
specific ranges of the parameters. Secondly, we have found two
particular reductions of $M_{37}$ - model by taking torsion and
curvature as independent cases. These reductions form $M_{44}$ -
$F(R)$ and $M_{45}$ - $F(T)$ theories of gravity inheriting general
functions $u$ and $v$ coming from the spacetime geometry. For two
viable DE models having specific properties, we have found some
constraints on constant parameters of the models to satisfy the
energy conditions.

One can use the effective gravitational field equations (\ref{9})
and (\ref{10}) to construct the expressions of energy conditions for
any viable model. The role of torsion depends upon the source
associated with spin matter density whereas in $F(R,T)$ gravity,
torsion as well as curvature can propagate without the presence of
spin matter density.

\end{document}